\documentclass[final,5p,times,twocolumn]{elsarticle}
\usepackage{amssymb}
\usepackage{amsmath}
\usepackage[utf8]{inputenc}
\usepackage{hyperref}
\usepackage{tikz}
\usepackage[T1]{fontenc}
\usepackage{algorithm}
\usepackage{algpseudocode}
\usepackage{multirow}
\usepackage{setspace}
\usetikzlibrary{shapes.geometric, arrows,calc}

\tikzstyle{1d} = [rectangle, rounded corners, minimum width=2.5cm, minimum height=0.7cm, text centered, draw=black, fill=green!40, text width=6.5cm]
\tikzstyle{2d} = [rectangle, rounded corners, minimum width=2.5cm, minimum height=0.7cm, text centered, draw=black, fill=blue!20, text width=6.5cm]
\tikzstyle{2dsep} = [rectangle, rounded corners, minimum width=2.5cm, minimum height=0.7cm, text centered, draw=black, fill=yellow!50, text width=6.5cm]
\tikzstyle{collective} = [rectangle, rounded corners, minimum width=2.5cm, minimum height=0.7cm, text centered, draw=black, fill=red!50, text width=6.5cm]
\tikzstyle{decision} = [diamond,aspect=2,minimum width=1cm, minimum height=0.7cm, text centered, draw=black, fill=red!50]
\tikzstyle{arrow} = [thick,->,>=stealth]

\renewcommand{\vec}{\boldsymbol}

\begin{document}
\begin{frontmatter}
\title{Three-dimensional Skyrme Hartree-Fock-Bogoliubov solver in coordinate-space representation}
\author[a,b]{Mengzhi Chen\corref{author}}
\ead{chenme24@msu.edu}
\author[a,b]{Tong Li}
\ead{lit@nscl.msu.edu}
\author[a]{Bastian Schuetrumpf}
\ead{bastian.schuetrumpf@gmail.com}
\author[c]{Paul-Gerhard Reinhard}
\ead{Paul-Gerhard.Reinhard@fau.de}
\author[a,b]{Witold Nazarewicz}
\ead{witek@frib.msu.edu}

\cortext[author] {Corresponding author}

\address[a]{Facility for Rare Isotope Beams, Michigan State University, East Lansing, Michigan 48824, USA}
\address[b]{Department of Physics and Astronomy, Michigan State University, East Lansing, Michigan 48824, USA}
\address[c]{Institut f{\"{u}}r Theoretische Physik, Universit{\"{a}}t Erlangen, D-91054 Erlangen, Germany}

\begin{abstract}
The coordinate-space representation of the Hartree-Fock-Bogoliubov  theory is the method of choice to study weakly bound nuclei whose properties are affected by  the quasiparticle continuum space.
To describe such systems, we developed a three-dimensional Skyrme-Hartree-Fock-Bogoliubov solver {\sc HFBFFT} based on the existing, highly optimized and parallelized Skyrme-Hartree-Fock code {\sc Sky3D}. The code does not impose any self-consistent spatial symmetries such as mirror inversions or parity. 
The underlying equations are solved in {\sc HFBFFT} directly in the canonical basis using the fast Fourier transform. 
To remedy the  problems with pairing collapse, we implemented the soft energy cutoff and pairing annealing. The convergence of HFB solutions was improved by a sub-iteration method.
The Hermiticity violation of differential operators  brought by Fourier-transform-based differentiation has also been solved.
The accuracy and performance of {\sc HFBFFT} were tested by benchmarking it against other HFB codes, both spherical and deformed, for a set of nuclei,  both well-bound and weakly-bound.
\end{abstract}

\begin{keyword}
Nuclear Density Functional Theory; Hartree-Fock-Bogoliubov method; Skyrme energy density functional; Finite volume methods; 3D coordinate-space representation
\end{keyword}
\end{frontmatter}

\section{Introduction}\label{sec:introduction}
Exotic nuclei with extreme neutron-to-proton ratios are crucial for  theoretical nuclear structure research as
their properties provide critical information on nuclear interactions, many-body techniques, and astrophysical scenarios. 
However, because of their weak binding, their quasiparticle excitations are often affected by  the low-lying scattering space (a.k.a.\ particle continuum), which enhances the necessary computational effort. For such nuclei, nucleonic pairing must be handled within the full Hartree-Fock-Bogoliubov (HFB) scheme instead of the simpler Bardeen-Cooper-Schrieffer (BCS) approximation \cite{Dobaczewski1984,Belyaev1987,Dobaczewski1996,Dobaczewski2013,Bulgac02}.  In addition, the associated self-consistent densities  are usually very extended in space, which requires large basis sets or large coordinate-space boxes.
Both requirements become particularly demanding if one aims at symmetry-unrestricted calculations (i.e., without imposing space reflection or axial or spherical symmetries). 
The paper aims to propose a reliable and efficient computational scheme  to solve the HFB equations  on a three-dimensional (3D) Cartesian coordinate-space grid. 

The nuclear energy density functional (EDF) method is one of the most widely used methods to study medium-mass and heavy nuclei \cite{Bender2003,Schunck2019}.
Its main ingredient  is an  EDF that represents the effective in-medium nuclear interaction.
Among many EDFs, the Skyrme functional, originally based on Skyrme interaction \cite{Skyrme1958}, is  commonly used to study global
 nuclear properties, such as ground-state energies, deformations, and low-lying excitations \cite{Schunck2019,Erler2011,Erler2012a}.
It is to be noted that the nuclear EDF method is closely related to the density-functional theory (DFT).
Hence, it is often referred to as nuclear DFT.

Over the years, a number of HFB solvers have been developed; see Table 2 of Ref.~\cite{Bender2020} for a summary. These solvers can be divided into two families.
The codes in the first group are based on the expansion of single-particle wave functions in a finite set of basis functions such as the harmonic
oscillator (HO) eigenfunctions. 
Examples of such solvers are: 
 {\sc HFBTHO} \cite{Stoitsov2005, Perez2017}, which  solves the axial (2D) HFB equations in the axial HO  or the transformed HO basis; 
{\sc HFODD} \cite{Dobaczewski1997_1, Schunck2017,Dobaczewski2021}, which  solves the 3D HFB equations in the  Cartesian HO basis without assumption of self-consistent symmetries; and {\sc HFBPTG}  \cite{Stoitsov2008}, which  solves the axial (2D) HFB equations in the P\"oschl-Teller-Ginocchio basis.

The basis-expansion method is efficient and has been successfully employed in large-scale calculations \cite{Erler2012a}.
However, when it is applied to  weakly-bound nuclei, the performance of this method deteriorates   as huge configuration spaces  are required to describe  the asymptotic behavior of HFB solutions. 
Here, the approach of choice is the HFB framework formulated in the coordinate-space representation \cite{Dobaczewski1984,Dobaczewski1996,Dobaczewski2013}.

The  coordinate-space solvers constitute the second family of HFB codes. Examples of such solvers are: 
 {\sc HFBRAD} \cite{Bennaceur2005} solves spherically symmetric HFB problem using finite differences; 
{\sc HFB-AX} \cite{Pei2008} is a 2D solver based on B-splines; 
{\sc SkyAx} \cite{Reinhard2020} is a highly optimized 2D  Hartree-Fock (HF) + BCS code using  the fast Fourier transform (FFT) method to compute derivatives;
{\sc Sky3D}  \cite{Maruhn2014,Schuetrumpf2018} is a 3D extension of {\sc SkyAx}; the predecessor of {\sc SkyAx} and {\sc Sky3D} is a 1D spherical HF+BCS code using five-point finite differences which was published first in \cite{Reinhard1991} and has meanwhile been developed into a full spherical  HFB code  {\sc Sky1D} \cite{PGHFBcodes};
the HFB extension of {\sc SkyAx} is {\sc Sky2D} \cite{PGHFBcodes};
{\sc EV8}  solves the Skyrme HF+BCS equations  using the imaginary time method on a 3D mesh that is limited to one octant  by imposing time-reversal and spatial symmetries \cite{Bonche2005,Ryssens2015}; {\sc MOCCa} \cite{Ryssens2019,Scamps2021,Scamps2021_2} is a Skyrme-HFB extension of {\sc EV8}; {\sc MADNESS-HFB} \cite{Pei2014} is a 3D HFB solver
based on multi-resolution analysis and multi-wavelet expansion; {\sc LISE} is a 3D HFB solver 
\cite{Jin2021} employing  the  discrete variable representation (or Lagrange-mesh method)
and fast Fourier transforms; and there are also 3D HFB solvers based on the contour integral of the Green's function using the shifted Krylov subspace method \cite{Jin2017,Kashiwaba2020}. 

The major difference between 
basis-based and mesh-based methods is the treatment of one-quasiparticle continuum space \cite{Belyaev1987,Michel2008,Pei2011,Dobaczewski2013}.
In the case of coordinate-space methods, the discretized continuum strongly depends on the geometry of the spatial box, the grid size, and the method employed for the representation of derivatives. 
For large 3D boxes and dense grids, the size of the discretized continuum space quickly becomes intractable as the maximum allowed quasiparticle energy increases.

A promising approach to the coordinate-space HFB problem is the canonical-basis HFB method  proposed  in Refs.\ \cite{Reinhard1997, Tajima2004}.
The one-body density matrix is diagonal in the canonical basis (or natural orbits), and its eigenstates are spatially localized if the nucleus is particle-bound. 
Because of this localization, the single-particle (s.p.) continuum level density  is significantly reduced.

In this work, we develop a  3D Skyrme-HFB solver {\sc HFBFFT} in the coordinate-space representation. This code is based on the published  code {\sc Sky3D} \cite{Maruhn2014,Schuetrumpf2018}.
{\sc Sky3D} has been well optimized for performance and parallelized with OpenMP and MPI \cite{Afibuzzaman2018}.
In {\sc HFBFFT} we maintain the high-level parallelization, making it scalable on modern supercomputers.
In order to overcome the pairing collapse problem mentioned in Ref.\ \cite{Tajima2004}, we implement the soft energy cutoff of pairing space and develop the annealing of pairing strengths to avoid pairing deadlock at an early stage.
Furthermore, we introduce the sub-iteration method in the configuration space to stabilize and speed up the convergence. 
We also resolve the problem of Hermiticity violation in {\sc Sky3D} brought by the incompatibility between the product rule and the Fourier-transform-based algorithm for derivatives.
To benchmark  {\sc HFBFFT} we study several nuclear systems and compare our results against {\sc HFBTHO} and the coordinate-space HFB codes {\sc Sky1D} and {\sc Sky2D}, which solve the HFB problem in 1D (spherical) and 2D (axial) geometries, respectively.

This paper is organized as follows.
In Sec.\ \ref{sec:basis}, the Skyrme EDF and the HFB theory are briefly introduced.
The numerical details and algorithms of {\sc HFBFFT} are described in section \ref{sec:algorithm}.
In Sec.\ \ref{sec:benchmark}, we present  test and benchmark results.
Finally, conclusions and outlook are presented in Sec.\ \ref{sec:conclusion}.

\section{Skyrme Hartree-Fock-Bogoliubov theory}\label{sec:basis}

In this section, we briefly summarize the Skyrme EDF and the HFB theory.

\subsection{The Skyrme energy density functional}

The HFB theory describes a many-Fermion system in terms
of an orthonormal set of s.p.\ wave functions $\psi_\alpha$
with fractional occupation amplitudes $v_\alpha$, i.e.,
\begin{equation}
    \left\{\psi_\alpha,v_\alpha,\alpha=1,...,\Omega\right\},
\label{eq:spbasis}
\end{equation}
where $\Omega$ denotes the size of the active s.p.\ space.  
The amplitude $v_\alpha$ can take values continuously in the interval $[0,1]$. The complementary  amplitude is $u_\alpha=\sqrt{1-v_\alpha^2}$. 

The code {\sc HFBFFT} uses a formulation of the HFB theory in the basis of natural orbitals, which are defined as the basis of s.p.\ states $\psi_\alpha$ in which the one-body density matrix $\hat{\rho}$ is diagonal, i.e.,
$\hat{\rho}=\sum_\alpha|\psi_\alpha\rangle{n}_\alpha\langle\psi_\alpha|$,
where $n_\alpha$,  an eigenvalue of $\hat{\rho}$, represents the canonical-state occupation.
The numerical HFB scheme in the canonical basis was presented in \cite{Reinhard1997} and improved
in \cite{Tajima2004}. For the relation between the standard matrix formulation and the canonical formulation of HFB, see Refs.~\cite{Dobaczewski1996,Ring2004}.
In the canonical basis, the HFB
mean-field state takes the BCS-like form:
\begin{equation}
  |\Phi\rangle
  =
  \prod_{\alpha>0}\big(
  u_\alpha^{\mbox{}}+v_\alpha^{\mbox{}}\hat{a}^+_\alpha\hat{a}^{+}_{\overline{\alpha}}
  \big)|0\rangle
\label{eq:BCState}
\end{equation}
where $|0\rangle$ is the vacuum state, $\hat{a}^+_\alpha$ is the
creation operator of $\psi_\alpha$, and $\overline{\alpha}$ the
conjugate partner to state $\alpha$ that corresponds to the same eigenvalue  of $\hat{\rho}$.

Any self-consistent mean-field theory starts from expressing the
energy of the system in terms of s.p.\ wave functions and occupation
amplitudes (\ref{eq:spbasis}).  
EDFs go for a simpler approach by starting from the energy defined in terms of only a few local densities and currents.
For the case of stationary states of even-even
nuclei, the energy depends only on the local particle density $\rho_q$, the kinetic-energy density
$\tau_q$, and the spin-orbit density $\vec{J}_q$:
\begin{subequations}\label{eq:densities}
\begin{align}
   \rho_q(\vec{r})&=\displaystyle
    \sum_{\alpha\in q}\sum_{s}  
    v_{\alpha}^2|\psi_{\alpha}(\vec{r},s)|^2,
   \notag \\
     \tau_q(\vec{r})&=\displaystyle
    \sum_{\alpha\in q}\sum_{s}  
    v_{\alpha}^2|\nabla\!\psi_{\alpha}(\vec{r},s)|^2, \notag \\
   \vec{J}_q(\vec{r}) &=\displaystyle
    -\mathrm{i}\sum_{\alpha\in q}\sum_{ss'} v_{\alpha}^2
    \psi_{\alpha}^*(\vec{r},s)
    \nabla\! \times\! \vec{\sigma}_{ss'} 
    \psi^{\mbox{}}_{\alpha}(\vec{r},s'),
 \label{eq:rtjeven}
\end{align}
where $q\in\{\mathrm{p},\mathrm{n}\}$ stands for protons or neutrons and $s, s'=\pm 1/2$ label the two spinor components of the wave functions.
Pairing EDFs additionally require the pairing density
\begin{align}
  \xi_q(\vec{r}) 
  &=\displaystyle
  {\sum_{\alpha\in q}u_{\alpha}v_{\alpha} 
    \sum_{s}\left( -2s \right) \psi_{\overline{\alpha}}(\vec{r},-s)
    \psi_{\alpha}}(\vec{r},s).
 \label{eq:rtjpair}
\end{align}
For a stationary state of an even-even nucleus, the conjugate s.p.\ state $\overline{\alpha}$ can be assumed to be the time-reversed state of $\alpha$, which leads to
\begin{align}
  \xi_q(\vec{r}) 
  &=\displaystyle
  {\sum_{\alpha\in q} \sum_{s} 
  u_{\alpha}v_{\alpha} \left| \psi_{\alpha}(\vec{r},s) \right|^2}.
\end{align}
\end{subequations} 

The code {\sc HFBFFT}, as its predecessor {\sc Sky3D}, employs the
widely used Skyrme EDF. This EDF is well described in all details at
several places  \cite{Bender2003,Erler2011,Schunck2019}.
Thus we give here only a brief account with emphasis on the pairing
part.  The total energy is a functional of the local densities:
\begin{subequations}
\label{eq:Etot}
\begin{equation}
    E_\mathrm{tot}
    =
    E_\mathrm{Skyrme}[\rho,\tau,\vec{J}]
    +
    E_\mathrm{pair}[\rho,\xi]
    +
    E_\mathrm{Coul}[\rho_\mathrm{p}],
\label{eq:efundet}
\end{equation}
where  (ignoring here isospin index for simplicity)
\begin{align}
	E_\mathrm{Skyrme} 
	&= E_\mathrm{kin} + E_\mathrm{\rho \rho} + E_\mathrm{\rho \tau} + E_{\rho \Delta \rho}
	+ E_\mathrm{ \nabla \vec{J}} 
\notag \\
   &= \int\! d^3r\left[ \frac{\hbar^2}{2m}\tau +
    C^{\rho}\rho^2 + C^{\tau}\rho \tau+ C^{\Delta\rho}\rho\Delta\rho + C^{\vec{J}}\rho \nabla\cdot\vec{J}\right],
   \\
  E_\mathrm{pair}
  &=
  \frac{1}{4} \sum_{q\in\{\mathrm{p},\mathrm{n}\}}V_{\mathrm{pair},q}
  \int d^3r |\xi_q|^2
  \left[1 -\frac{\rho}{\rho_{0,\mathrm{pair}}}\right],
  \\
  E_{\mathrm{Coul}}
  &=\frac{e^{2}}{2} \int\! \mathrm{d}^{3} r \mathrm{~d}^{3} r^{\prime} \frac{\rho_{\mathrm{p}}(\vec{r}) \rho_{\mathrm{p}}\left(\vec{r}^{\prime}\right)}{\left|\vec{r}-\vec{r}^{\prime}\right|}
  -\int\! \mathrm{d}^{3} r \frac{3 e^{2}}{4}\left(\frac{3}{\pi}\right)^{\frac{1}{3}} \rho_{\mathrm{p}}^{4 / 3}.
\label{eq:epair}
\end{align}     
\end{subequations}
 $E_\mathrm{Skryme}$ is a functional of $\rho$, $\tau$,
and $\vec{J}$; $E_\mathrm{pair}$ is a functional of $\rho$
and $\xi$;
and the Coulomb energy $E_{\mathrm{Coul}}$ is a functional of proton density $\rho_\mathrm{p}$. 
The pairing functional can be motivated by a density-dependent
$\delta$ interaction. 
It includes two limiting cases.  
The first case is a pure  contact interaction, also called volume pairing, which is recovered when $\rho_{0,\mathrm{pair}}\rightarrow\infty$. 
The second case corresponds to a value near the equilibrium density $\rho_{0,\mathrm{pair}}=0.16$ fm$^{-3}$, which localizes pairing around the nuclear surface.  
Adjustment of $\rho_{0,\mathrm{pair}}$ as a free parameter delivers a form of the pairing functional which stays in between the extremes of volume and surface pairing \cite{Dobaczewski2001,Kluepfel2009}.

\subsection{The HFB theory in canonical basis} \label{sec:HFB_theory}

In practice, one deals with two types of fermions: protons and  neutrons. 
To keep the notation simple, in the following, we assume that  the isospin quantum number is included in the quantum label $\alpha$ of the canonical state. 
The HFB equations are derived variationally  by minimizing the HFB Routhian
\begin{equation}\label{routhian}
    R=E_{\mathrm{tot}}
    -\sum_{q\in \{\mathrm{p},\mathrm{n}\}}\epsilon_{\mathrm{F},q} \sum_{\alpha\in q} v_{\alpha}^{2}
    - \sum_{\alpha\beta} \lambda_{\alpha\beta}\left(\langle\psi_{\beta} | \psi_{\alpha}\rangle-\delta_{\alpha\beta}\right),
\end{equation}
with respect to $\psi_\alpha$ and $v_\alpha$.
In Eq.~(\ref{routhian})
$\epsilon_\mathrm{F}$ is the
Fermi energy which is also the Lagrange parameter for the particle-number constraint, 
and the $\hat\lambda$ is the matrix of Lagrangian
multipliers that guarantee the  orthonormality of canonical wave functions. 
Since $\langle\psi_{\beta} | \psi_{\alpha}\rangle = \langle\psi_{\alpha} | \psi_{\beta}\rangle^*$,
it is required that the matrix $\hat\lambda$
is Hermitian  so that the number of its independent elements coincides with the total number of independent constraints.

Variation of the Skryme and Coulomb energies with regard to the s.p.\ wave function yields the HF Hamiltonian
$\hat{h}$:
\begin{equation}
  \frac{\delta \left(E_\mathrm{Skyrme} + E_\mathrm{Coul} \right)}{\delta\psi_\alpha^\dagger}
  =
  v_\alpha^2\hat{h}\psi_\alpha.
\label{eq:mfham}
\end{equation}
By the chain rule for derivatives, (\ref{eq:mfham}) can be reduced to the variation with respect to the densities, which delivers explicit expressions for $\hat{h}$ \cite{Bender2003,Erler2011}.
The HF Hamiltonian $\hat{h}$ is a functional of local densities (particle density, kinetic-energy density, spin-orbit density) in the standard fashion of nuclear EDFs \cite{Bender2003}. 

Variation of the pairing energy with respect to the s.p.\ wave function gives
\begin{equation}\label{eq:pair_ham}
    \frac{\delta E_\mathrm{pair}}{\delta\psi_\alpha^\dagger}
    =
    u_\alpha v_\alpha \hat{\tilde{h}}\psi_\alpha + v_\alpha^2 \hat{h}^\prime \psi_\alpha.
\end{equation}
The first term is related to the  variation
with respect to the pairing density, which yields the pairing potential \cite{Ring2004}
\begin{equation}
  \tilde{h}_q(\vec{r})
  =
  \frac{1}{2} V_{\mathrm{pair},q}
  \xi_q
  \left[1 -\frac{\rho}{\rho_{0,\mathrm{pair}}}\right],\ q\in \{\mathrm{p},\mathrm{n}\}.
\label{eq:gappot}
\end{equation}
The second term is the pairing-rearrangement term, brought by the density dependence of the pairing functional. 
For simplicity, we treat the rearrangement term $\hat{h}^\prime$ as part of the HF Hamiltonian $\hat{h}$ in the following. 
The pairing potential $\tilde{h}_q(\vec{r})$ is a local potential in most pairing functionals.
From that, we obtain the state-dependent gap
\begin{equation}
  \Delta_{\alpha\alpha}
  =
  \left|\langle\psi_\alpha|
  \tilde{h}_{q_\alpha}|\psi_\alpha\rangle\right|
  ,
\label{eq:gapalpha}
\end{equation}
where $q_\alpha$ is the isospin of state $\alpha$. 
Another aspect of the pairing is determined by the gap equation, which is obtained from the variation with respect to  $v_\alpha$:  
\begin{equation}
  0
  =
  4 v_\alpha^{\mbox{}} (h_{\alpha\alpha}^{\mbox{}} - \epsilon_{\mathrm{F},q_\alpha})
  + 2 \left(\frac{v_\alpha^2}{u_\alpha^{\mbox{}}}-u_\alpha^{\mbox{}}\right)
   \Delta_{\alpha\alpha},
\label{eq:gapeq}
\end{equation}
where $h_{\alpha\alpha}$ are the diagonal matrix elements of
the HF Hamiltonian $\hat{h}$.
The HF Hamiltonian together with the pairing potential constitutes the main ingredients of the HFB equations. 

With the orthonormality of canonical states taken into account, the constrained variation of the total energy with respect to $\psi_\alpha^\dagger$ yields the mean-field equations:
\begin{equation}
  \hat{\mathcal{H}}_\alpha^{\mbox{}} \psi_\alpha^{\mbox{}}
=
  \textstyle{\sum_\beta}
  \psi_\beta\lambda^{\mbox{}}_{\beta\alpha}, 
\label{eq:cmfeq}
\end{equation}
where
\begin{subequations}\label{eq:genmf_avlambda}
\begin{align}
  \hat{\mathcal{H}}_\alpha^{\mbox{}} 
  &=
  v^2_\alpha \hat{h} + u_\alpha^{\mbox{}}
  v_\alpha^{\mbox{}} \hat{\tilde{h}},
\label{eq:genmf}
\\
  \lambda^{\mbox{}}_{\beta\alpha}
  &=
  \frac{1}{2}
  \langle\psi_\beta|\hat{\mathcal{H}}_\alpha^{\mbox{}}+\hat{\mathcal{H}}_\beta|\psi_\alpha^{\mbox{}}\rangle.
\label{eq:avlambda}
\end{align}
\end{subequations}
The mean-field equations (\ref{eq:cmfeq},\ref{eq:genmf_avlambda}) and  gap equations (\ref{eq:gapeq}) together constitute the self-consistent HFB equations in the canonical basis. 

In (\ref{eq:genmf})  $\hat{\mathcal{H}}_\alpha^{\mbox{}}$ is a state-dependent one-body Hamiltonian composed of the HF Hamiltonian and the pairing potential.
The full matrix $\hat\lambda$ needs to be taken into account because the $\hat{\mathcal{H}}_\alpha^{\mbox{}}$ is state-dependent \cite{Reinhard1997,Tajima2004}.
In contrast, pure HF or HF+BCS calculations only require diagonal matrix elements $\lambda_{\alpha\alpha}$, which are also known as s.p.\ energies. 
The Hermiticity of  $\hat\lambda$ is enforced by explicit symmetrization in Eq.\ (\ref{eq:avlambda}).
It can be shown by multiplying both sides of Eq.\ (\ref{eq:cmfeq})   by  $\psi_\beta^\dagger$ that the final solution should obey the symmetry conditions
\begin{subequations}
\label{eq:symmcond_both}
\begin{equation}
   0
   =\lambda^{-}_{\beta\alpha}\equiv
   \frac{1}{2}\left(\langle\psi_\beta^{\mbox{}}|\hat{\mathcal{H}}_\alpha^{\mbox{}}|\psi_\alpha^{\mbox{}}\rangle
   -
   \langle\psi_\beta^{\mbox{}}|\hat{\mathcal{H}}_\beta^{\mbox{}}|\psi_\alpha^{\mbox{}}\rangle\right).
\label{eq:symmcond}
\end{equation}
One can combine these into one condition:
\begin{equation}
   0
   =
   \Delta\mathcal{S}
   \equiv\frac{1}{2}\sum_{q\in\{\mathrm{p,n}\}}\sqrt{
   \frac{1}{\Omega_q^2} \sum_{\alpha, \beta \in q} \left| \lambda^{-}_{\beta\alpha} \right|^2}. 
\label{eq:symmcondaver}
\end{equation}
\end{subequations}
The actual size of $\Delta\mathcal{S}$ will serve as a check for the convergence of the HFB solution.

It should be noted that $\lambda^-_{\beta\alpha}$ vanishes when both s.p.\ states $\psi_\alpha$ and $\psi_\beta$ are fully occupied ($v_\alpha = v_\beta = 1$) or unoccupied ($v_\alpha = v_\beta = 0$) since then $\langle\psi_\alpha|\hat{h}|\psi_\beta\rangle=\langle\psi_\beta|\hat{h}|\psi_\alpha\rangle^*$. 
Thus, for a pure HF calculation, $\Delta\mathcal{S}$ measures the overlap between occupied and unoccupied orbits, which should be zero at the self-consistent solution.

When the size of active s.p.\ space equals the number of particles ($\Omega_\mathrm{n} = N$, $\Omega_\mathrm{p} = Z$) and 
all the s.p.\ orbits are fully occupied (as in the pure HF calculation), 
$\Delta\mathcal{S}$ is always  zero; hence, it is not an appropriate measure for convergence.
However, this quantity can still be utilized to check the Hermiticity of our implementation of $\hat{h}$ (see Sec.\ \ref{sec:hermiticity}). 
For the pure HF case, a suitable quantity for convergence check could be
\begin{equation}
    \sqrt{\sum_\alpha \left|\langle \psi_\alpha | \hat{h}^2 | \psi_\alpha \rangle - \langle \psi_\alpha | \hat{h} | \psi_\alpha \rangle^2\right|},
\end{equation}
which approaches zero for the converged HF solution.

\section{Numerical representation} \label{sec:algorithm}

\subsection{Numerical realization on a 3D coordinate-space grid} \label{sec:num_grid}

The numerical representation is explained in detail in Refs.~\cite{Maruhn2014,Schuetrumpf2018}. 
Here, we repeat the essentials briefly. For simplicity, our discretization strategy is explained here for one dimension;
the generalization to 3D is straightforward.

All wave functions, densities and fields are defined on a three-dimensional equidistant Cartesian grid. 
The grid points in the $x$ direction are
\begin{equation}
  x_\nu = \left(-\frac{N_x+1}{2}+\nu\right)\delta x,
  \quad \nu=1,\ldots,N_x,
\label{eq:xgrid}
\end{equation}
where $N_x$ is the (even) number of grid points and $\delta x$ is the
grid spacing. 
Similar gridding applies to the $y$ and $z$ directions.  The action of local
operators on a coordinate-space grid is a simple multiplication
of the local operator field and the wave function.  
The action of momentum operators, such as in the kinetic energy, requires first and second derivatives defined  in Fourier space.
The Fourier technique has been proved to be superior in precision and advantageous for large grids \cite{Blum1992}.
It is noteworthy that the direct Coulomb potential is also solved in Fourier space.
The Coulomb solver has to fulfill the condition that the result in the  box is the correct solution to Poisson's equation with the boundary condition of zero potential at infinity.
The algorithm to solve Poisson's equation for an isolated charged distribution has been implemented in {\sc Sky3D}.
It follows the ideas of \cite{Hockney1970, Eastwood1979} by doubling the 3D grid, folding the proton density with the $1/r$ Green's function in momentum space and then restricting the final solution inside the original box.
A subtle point is the setting of the Coulomb field at $k=0$ in reciprocal  space. We optimized it empirically to typical nuclear situations, for details see \cite{Maruhn2014}, which leaves a possible uncertainty of a few keV just at the edge of the last digit in the comparisons presented in this paper.

The discrete grid points $k_n$ in Fourier space are related to the same number of grid points $x_\nu$ in coordinate space as:
\begin{subequations}
\label{eq:FT}
\begin{align}
k_n  &=\left\{
\begin{aligned}
 &(n-1)\delta k, \quad n=1,\ldots, \frac{N_x}{2} \\
 &(n-N_x-1)\delta k, \quad n=
   \frac{N_x}{2}+1,\ldots, N_x 
\end{aligned},
\right. \\
  \delta k
  &=
  \frac{2\pi}{N_x\delta x}.
\end{align}
\end{subequations}
Note that the coordinate-space grid (\ref{eq:xgrid}) in combination with the conjugate momentum-space grid (\ref{eq:FT}), imposes no spatial symmetry at all. 
But the particular examples considered for benchmarking in this study obey reflection symmetry in all three directions.

A wave function $\psi(x_\nu)$ in coordinate space is related to a
wave function $\widetilde{\psi}(k_n)$ in Fourier space by the discrete Fourier transform and its inverse
\begin{subequations}
\begin{align}
  \widetilde{\psi}(k_n)
  &=
  \sum_{\nu=1}^{N_x}
  \exp{\left(-\mathrm{i} k_nx_\nu\right)}\psi(x_\nu) 
  ,
\label{eq:FTforward}\\
  \psi(x_\nu) 
  &=
  \frac{1}{N_x}\sum_{n=1}^{N_x}
  \exp{\left(\mathrm{i} k_nx_\nu\right)}\widetilde{\psi}(k_n).
\label{eq:FTbackward}
\end{align}
\end{subequations}
Both can be efficiently computed via the FFT algorithm provided by the FFTW3 library \cite{Frigo2005}.
This complex Fourier representation implies that the function $\psi$ is
periodic, i.e., $\psi(x+N_x \cdot \delta x)=\psi(x)$. The appropriate
integration scheme that complies with the above summations is the trapezoidal rule
\begin{equation}
    \int_{-\frac{N_x}{2}\delta x}^{\frac{N_x}{2}\delta x} dx\, f(x) \approx \sum_{\nu = 1}^{N_x} f(x_\nu) \delta x, 
\end{equation}
where all terms are added up with equal weights.

In Fourier space the $m$-th
derivative becomes a multiplication by $(\mathrm{i} k_n)^m$. 
One proceeds then in the following way: First, a forward transform (\ref{eq:FTforward}) is performed; then  $\widetilde{\psi}(k_n)$ is multiplied by
$(\mathrm{i} k_n)^m$; and finally 
$(\mathrm{i}k_n)^m\widetilde{\psi}(k_n)$ is transformed back to the coordinate space by Eq.~(\ref{eq:FTbackward}).  
One should note that there is an arbitrariness about the choice of momentum $k_{N_x/2+1}$:
it can be taken as $\pm \frac{N_x}{2} \delta k$.
This arbitrariness does not alter the transforms (\ref{eq:FTforward}, \ref{eq:FTbackward}) 
but gives different results of the $m$-th derivative when $m$ is odd (no impact when $m$ is even).
A natural choice is to equally split $\widetilde{\psi}(k_{N_x/2+1})$ between the positive and negative momenta,
making them cancel each other in the final result of an odd-order derivative.
It is equivalent to setting  $\widetilde{\psi}(k_{N_x/2+1})=0$.
This choice ensures that the derivative of a real-valued function is still real-valued; 
it also means that the second derivative is not equivalent to two consecutive first derivatives in this framework.
The remaining problem is the Hermiticity breaking caused by the product rule;
we will discuss it in Sec. \ref{sec:hermiticity}.

\subsection{Solution by accelerated gradient iteration}
\label{sec:grad}

The solution of the coupled HFB equations is obtained by interlaced iterations of the gap equation and the mean-field equation. 
The gap equation (\ref{eq:gapeq}) can be solved in a closed form and it yields:
\begin{equation}
  \left\{\begin{array}{c} v_\alpha^{\mbox{}} \\ u_\alpha^{\mbox{}} \end{array}\right\}
  =
  \sqrt{\frac{1}{2}\mp \frac{1}{2}
      \frac{h_{\alpha\alpha}^{\mbox{}}-\epsilon_{\mathrm{F},q_\alpha}}
           {\sqrt{(h_{\alpha\alpha}-\epsilon_{\mathrm{F},q_\alpha})^2+\Delta^2_{\alpha\alpha}} } }
   .
\label{eq:uveq}
\end{equation}
The Fermi energy $\epsilon_\mathrm{F}$ needs to be adjusted to fulfill the particle-number condition
\begin{equation}
  \epsilon_{\mathrm{F},q}\;\longleftrightarrow\;
  \sum_{\alpha\in q}^{\mbox{}} v_\alpha^2=N_q,
\end{equation}
where $N_q$ is the required particle number.
Note that only the diagonal elements of the pairing potential and the HF Hamiltonian in the canonical basis enter (\ref{eq:uveq}); hence, no information about the non-diagonal elements is needed to determine the occupation amplitudes.

The solution of the mean-field equation (\ref{eq:cmfeq}) is obtained by the damped gradient iteration \cite{Reinhard1982,Bottcher1989,Blum1992,Maruhn2014} interlaced with updating the matrix $\hat\lambda$ for the orthonormality constraint. 
The steps are:
\begin{enumerate}
  \item
    For given
    $\{\psi_\alpha^{\mbox{}},v_\alpha^{\mbox{}},u_\alpha^{\mbox{}},\alpha=1,...,\Omega\}$
    compute the local densities, the HF Hamiltonian $\hat{h}$ and the pairing potential $\hat{\tilde{h}}$.
  \item\label{it:hmf}
    Compute the action of $\hat{h}$ and $\hat{\tilde{h}}$ on all $\psi_\alpha^{\mbox{}}$
    and store the result in work arrays $\Psi_{\alpha}$ and $\widetilde{\Psi}_{\alpha}$, i.e.,
    \begin{subequations}
    \begin{equation}
      \hat{h}\psi_\alpha^{\mbox{}}\longrightarrow\Psi_\alpha^{\mbox{}},
    \end{equation}
    \begin{equation}
     \hat{\tilde{h}}\psi_\alpha^{\mbox{}}\longrightarrow\widetilde{\Psi}_\alpha^{\mbox{}},
    \end{equation}
    \end{subequations}
    for $\alpha=1,\ldots,\Omega$.
  \item\label{it:h_Delta_diag}
    Use  $\Psi_\alpha^{\mbox{}}$ and $\widetilde{\Psi}_\alpha^{\mbox{}}$ to compute and store the 
    s.p.\ energies and pairing gaps
    \begin{subequations}
    \begin{equation}
      h_{\alpha\alpha}=\langle\psi_\alpha^{\mbox{}}|\Psi_\alpha^{\mbox{}}\rangle,
    \end{equation}
    \begin{equation}
        \Delta_{\alpha\alpha} = \left| \langle\psi_\alpha^{\mbox{}}|\widetilde{\Psi}_\alpha^{\mbox{}}\rangle \right|.
    \end{equation}
    \end{subequations}
  \item
    Evaluate and store the action of the generalized mean-field Hamiltonian (overwriting $\Psi_\alpha^{\mbox{}}$)
    \begin{equation}
      \mathcal{H}_\alpha^{\mbox{}}\psi_\alpha^{\mbox{}}
      =
      v_\alpha^2\Psi_\alpha^{\mbox{}}+u_\alpha^{\mbox{}} v_\alpha^{\mbox{}}\widetilde{\Psi}_\alpha^{\mbox{}}
      \;\longrightarrow\;
      \Psi_\alpha^{\mbox{}}.
    \end{equation}
  \item
    Apply the matrix of Lagrange multipliers on all $\psi_\alpha$; compute and store (again overwriting $\Psi_\alpha^{\mbox{}}$)
    \begin{equation}
      \Psi_\alpha^{\mbox{}}-\sum_\beta^{\mbox{}}\psi_\beta^{\mbox{}}\lambda_{\beta\alpha}^{\mbox{}}
      \;\longrightarrow\;
      \Psi_\alpha^{\mbox{}}
      .
    \end{equation}
  \item
    Apply the damping operation $\hat{\mathcal{D}}$ and orthonormalization $\hat{\mathcal{O}}$
    \begin{subequations}
    \begin{equation}
      \psi_\alpha^\mathrm{(new)}
     =
      \hat{\mathcal{O}}\left\{\psi_\alpha^{\mbox{}}-\hat{\mathcal{D}}\Psi_\alpha^{\mbox{}}\right\}, 
      \end{equation}
      
        \begin{equation}
      \hat{\mathcal{D}}
      =
      \frac{x_0}{v_\alpha^2(\hat{T}+E_0)+\frac{1}{2}u_\alpha^{\mbox{}} v_\alpha^{\mbox{}}\tilde{h}_0}
      ,
    \end{equation}
    \end{subequations}
    where $x_0$, $E_0$, and $\tilde{h}_0$ are adjustable numerical parameters. 
    The empirical values $x_0=0.45$, $E_0=100$ MeV and $\tilde{h}_0=\mathrm{max}\left[\tilde{h}_\mathrm{n}(\vec{r}),\tilde{h}_\mathrm{p}(\vec{r})\right]$ are used in our calculations.
    It is worth noting that  the lower bound of $u_\alpha v_\alpha$ and $ v_\alpha^2$  in $ \hat{\mathcal{D}}$  is set to be $10^{-1}$ for numerical stability.
        
   \item\label{it:pair}
    With the new $h_{\alpha\alpha}$ and $\Delta_{\alpha\alpha}$ from step \ref{it:h_Delta_diag}, compute
    new occupations $v_\alpha^{\mbox{}}$ and $u_\alpha^{\mbox{}}$ 
    using Eq.~(\ref{eq:uveq}).
    
    \item Reevaluate the action of the generalized mean-field Hamiltonian on all $\psi_\alpha$ and compute the matrix of Lagrange multipliers 
    \begin{equation}
        \lambda^{\mbox{}}_{\beta\alpha}
      =
      \frac{
      \langle\psi_\beta|\hat{\mathcal{H}}_\alpha^{\mbox{}}|\psi_\alpha^{\mbox{}}\rangle 
      + \langle \psi_\alpha | \hat{\mathcal{H}}_\beta | \psi_\beta \rangle^*
      }{2}.
    \end{equation}
\end{enumerate}

The above iteration usually starts from a number of HF+BCS steps, which are done in the same way as in {\sc Sky3D}. 
The HF+BCS calculation is initialized by a 3D HO wave function that can be triaxially deformed.
To achieve better convergence, in step 1 the new densities are mixed linearly with the old ones:
\begin{equation}
    \kappa^{(n)} = (1-\gamma)\kappa^{(n-1)} + \gamma\kappa^{(n)}_{\psi}, \quad \kappa = \rho, \tau\ \mathrm{or}\ \xi,
\end{equation}
where $n$ is the iteration number, subscript $\psi$ denotes the density directly computed from the wave functions, and $\gamma$ is the adjustable mixture ratio with a default value of 0.2.

\subsection{Sub-iterations in configuration space}

The damped gradient scheme outlined in Sec.\ \ref{sec:grad} converges, but requires more iterations in the HFB scheme as compared to the HF + BCS used  in {\sc Sky3D}.
It also involves operations on the full 3D grid which  can make computations cumbersome.  
The pairing part in the iterative steps works predominantly within the given space of canonical  states.
Thus one can reduce the total numerical expense by the sub-iteration method: 
switching between the full 3D step and a fast iterative solver in configuration space. 
To this end, we map the mean-field equations into configuration space with
the expansion
\begin{equation}
  \psi_\alpha
  =
  \sum_{n=1}^{\Omega}\varphi_n c_{n\alpha},
  \label{eq:config_expansion}
\end{equation}
where $\{\varphi_n\}$ is a set of  s.p.\ states acting as the expansion
basis.  
For simplicity we choose an expansion basis such that $c^{(0)}_{n\alpha} = \delta_{n\alpha}$ at the beginning.
Inserting (\ref{eq:config_expansion}) into the HFB mean-field equations
(\ref{eq:cmfeq}) yields
\begin{equation}
  \lambda^-_{\beta \alpha} =
  \sum_{m n} c_{n\beta}^*
  \left\langle\varphi_n\left|\frac{\hat{\mathcal{H}}_\alpha^{\mbox{}}
                  -\hat{\mathcal{H}}_\beta^{\mbox{}}}{2}
 \right|\varphi_m\right\rangle 
  c_{m\alpha} = 0.
\label{eq:asymm}
\end{equation}
Eq.\ (\ref{eq:asymm}) is essentially the same as the symmetry condition (\ref{eq:symmcond}).
It is solved by a simple damped gradient iteration:
\begin{equation}
\begin{split}
  c_{n\alpha}^\mathrm{(new)}
  &=
\hat{\mathcal{O}}\left\{c_{n\alpha}^{\mbox{}}
  -
  \frac{\delta}{h_{nn}\!-\!h_{11}\!+\!E_0}
  \left[\sum_{m}\mathcal{H}_{\alpha,nm}^{\mbox{}} c_{m\alpha}^{\mbox{}}
       -\sum_{\beta}c_{n\beta}^{\mbox{}}\lambda_{\beta\alpha}^{\mbox{}}
  \right]
  \right\}
  \\
  &=\hat{\mathcal{O}}\left\{c_{n\alpha}^{\mbox{}}
  -
  \frac{\delta}{h_{nn}\!-\!h_{11}\!+\!E_0}
  \sum_{\beta}c_{n\beta}^{\mbox{}}\lambda^-_{\beta\alpha}
  \right\}
  ,
\end{split}
\label{eq:symm}
\end{equation}
where
$\mathcal{H}_{\alpha,nm}=\langle\varphi_n|\hat{\mathcal{H}}_\alpha^{\mbox{}}|\varphi_m\rangle$ and
\begin{equation}
\lambda_{\beta \alpha} = \frac{1}{2} \sum_{mn} c_{n\beta}^* \left( \mathcal{H}_{\alpha,nm} + \mathcal{H}_{\beta,nm} \right) c_{m\alpha}.
\end{equation}
The (interlaced) solution of the gap equations remains as before, but we do not update the local densities, the HF Hamiltonian $\hat{h}$ and the pairing potential $\hat{\tilde{h}}$ in configuration space.  
The convergence of the iteration is checked,
again, by the symmetry condition (\ref{eq:symmcondaver}).  
The most efficient combination of the full 3D step with the iterations in configuration
space is a matter of experience, see Sec.\ \ref{sec:benchmark}.

\subsection{Soft cutoff on pairing-active space}
It is well known that the HFB equations with local interactions diverge when solved in infinite quasiparticle/canonical space \cite{Dobaczewski1984}.  
To limit the pairing-active space,  all local densities (\ref{eq:densities}) are augmented by the cutoff factor $w_\alpha$, for instance the particle and pairing densities:
\begin{subequations}
\label{eq:cutpairdens}
\begin{align}
	\rho(\vec{r}) &= \displaystyle
    \sum_{\alpha}  
    w_{\alpha}v_{\alpha}^2\sum_{s}|\psi_{\alpha}(\vec{r},s)|^2,
    \\
  \xi(\vec{r}) &=
  \sum_{\alpha}w_{\alpha}^{\mbox{}}u_{\alpha}^{\mbox{}}v_{\alpha}^{\mbox{}}
  \sum_{s}|
   \psi_{\alpha}^{\mbox{}}(\vec{r},s) |^2.
\label{eq:pairdens}
\end{align}
The same augment also applies to the kinetic-energy and the spin-orbit densities.
A fixed number of states (realized by setting $w_\alpha=1$ or 0) is dangerous for two reasons.  
First, it hinders the portability of the pairing functional between codes and nuclei, because the s.p.\ space depends on the basis representation.
Second, level crossings near the hard cutoff can induce jumps of the pairing energy.
These problems can be solved by pairing renormalization \cite{Dobaczewski1996,Borycki2006} which, however, could be impractical in a full 3D treatment that involves huge canonical spaces.
Therefore, a commonly used remedy is to use a soft pairing cutoff \cite{Bonche1985,Krieger1990}
\begin{equation}
  w_\alpha^{\mbox{}}
  =
  \frac{1}
       {\displaystyle 1+
        \exp\left(\frac{h_{\alpha\alpha}-\epsilon_\mathrm{F}-\Delta\epsilon_\mathrm{cut}}
                      {\Delta\epsilon_\mathrm{cut}/10}
        \right)}
  .
\label{eq:softcut}
\end{equation}
\end{subequations}
The cutoff places a fixed band $\Delta\epsilon_\mathrm{cut}$ above
the actual Fermi energy $\epsilon_\mathrm{F}$.  We are going to use here $\Delta\epsilon_\mathrm{cut}=15$ MeV. 
It is  important to note that the soft cutoff modifies the state-dependent Hamiltonian $\hat{\mathcal{H}}_\alpha^{\mbox{}}$:
\begin{equation}
  \hat{\mathcal{H}}_\alpha^{\mbox{}} = w_\alpha \left( v^2_\alpha \hat{h} + u_\alpha^{\mbox{}}v_\alpha^{\mbox{}} \hat{\tilde{h}} \right),
\end{equation}
which defines all the ingredients entering the canonical HFB equations.

\subsection{Strategies to avoid premature pairing breakdown}
\label{sec:breakdown}

The pairing comes along with a second-order superfluid-to-normal phase transition.  Below the  critical pairing strength, the HFB pairing gap remains exactly zero. 
Above this critical strength, pairing becomes active and the gap starts to grow quickly. 
However, the onset of pairing is often delayed in a numerical calculation. 
The problem is that zero pairing remains a valid solution to the HFB (BCS) equations, but an unstable one.
It can then take a very long time before the algorithm overcomes the instability and drives towards a stable solution. 
As a consequence, an iteration scheme can easily be deadlocked due to a pairing breakdown. 
This is a well-known problem.
Most algorithms incorporate recovery strategies, such as occasional kickoffs by giving the pairing gap an artificial value, small enough not to spoil the physics but large enough to revive the pairing mechanism.

There is a more insidious problem with the state-dependent pairing gap $\Delta_{\alpha\alpha}$: 
it can happen that one canonical state logs out from the
pairing scenario and gets stuck in its own pairing breakdown
$\Delta_{\alpha\alpha}\rightarrow 0$. 
To understand that, we inspect  Eq.\ (\ref{eq:cmfeq}) and recall that 
$\hat{\mathcal{H}}_\alpha^{\mbox{}}=v_\alpha^{\mbox{}}\left(
v_\alpha^{\mbox{}}\hat{h} + u_\alpha^{\mbox{}}\hat{\tilde{h}}\right)$.
Far above the Fermi energy, we encounter states with $u_\alpha^{\mbox{}}\gg v_\alpha^{\mbox{}}$ such that $\hat{\mathcal{H}}_\alpha^{\mbox{}}\approx\hat{\tilde{h}}$ becomes a purely local operator. 
The solution to the mean field equation is $\psi\propto\delta(\vec{r}-\vec{r}_\mathrm{min})$ where $\vec{r}_\mathrm{min}$ is the point $\hat{\tilde{h}}$ has a minimum. 
In practice, this will be the representative of a
$\delta$-function on the grid, slightly mellowed by orthonormalization to other states.
As a consequence, the state acquires a very high kinetic energy and a very high canonical s.p.\ energy, which drives the solution of the gap equation (\ref{eq:uveq}) even more toward $v_\alpha^{\mbox{}}\rightarrow 0$. 
This as such is a valid physical mechanism as long as the iterations curb down the occupations slowly from above.
It becomes a
problem if some $v_\alpha^{\mbox{}}$ gets stuck at zero at the very early stage of the iterative process. 
Once this has happened, the state $\alpha$ is locked out of the pairing space.
In order to avoid this from happening, we adopt a strategy similar to simulated annealing \cite{Press1992} and start the iteration scheme with an enhanced effective pairing strength which gradually reduces to the
physical strength as
\begin{equation}
  V_\mathrm{pair}^\mathrm{(eff)}
  =  
  V_\mathrm{pair}^{\mbox{}}
  \left(\eta_\mathrm{enh}
    \frac{\mbox{max}(\mathcal{N}_\mathrm{enh}-\mbox{\tt iter},0)}
         {\mathcal{N}_\mathrm{enh}}
    +1
  \right),
\end{equation}
where {\tt iter} is the iteration number. 
In practice, we use an enhancement factor $\eta_\mathrm{enh}=2$ and $\mathcal{N}_\mathrm{enh}=400$. 
With this choice, the lock-in problem in the most critical early phases of iterations is avoided.

\subsection{Hermiticity restoration} \label{sec:hermiticity}
According to Refs.\ \cite{Maruhn2014,Schuetrumpf2018},
the explicit expression of applying the Skyrme HF Hamiltonian $\hat{h}$ on a wave function $\psi$ can be written as:
\begin{equation}\label{eq:h_terms}
	\begin{split}
	\hat{h}\psi =\ &U(\vec{r})\psi - \nabla\cdot \left[B(\vec{r})\nabla\right]\psi \\
     + & \frac{\mathrm{i}}{2}\left[\vec{W} \cdot\left(\vec{\sigma} \times \nabla\right)\psi + \vec{\sigma} \cdot \nabla \times \left(\vec{W}\psi\right) \right]. 
	 \end{split}
\end{equation}
This expression can be directly derived from the Skyrme EDF via Eq.\ (\ref{eq:mfham}), without invoking the product rule.
In \cite{Schuetrumpf2018} it was noted that the product rule is not perfectly fulfilled
when derivatives are evaluated via the discrete Fourier transform.
Therefore, in {\sc Sky3D} version 1.1 the commonly-adopted form of the spin-orbit term 
\begin{equation} \label{eq:old_spin_orbit}
\mathrm{i} \vec{W} \cdot(\vec{\sigma} \times \nabla) \psi
\end{equation}
was replaced by the one given in Eq.\ (\ref{eq:h_terms});
with $\nabla \times \vec{W} = 0$, these two forms are connected by the product rule.
However, the second term of $\hat{h}\psi$, which involves a position-varying differential operator, is still calculated through the product rule in {\sc Sky3D}:
\begin{equation}
  \nabla\cdot \left[B(\vec{r})\nabla\right]\psi = 
  \sum_{i=x,y,z} \frac{\partial B}{\partial i}\frac{\partial \psi}{\partial i} + B\frac{\partial^2 \psi}{\partial i^2}.
  \label{eq:prodrule}
\end{equation}
Unfortunately,  evaluating Eq.\ (\ref{eq:prodrule}) with the FFT-based differentiation breaks the Hermiticity of the operator \cite{Johnson2011}.
This point is confirmed by computed results shown in Sec.\ \ref{sec:hermiticityresult}.

Instead of using Eq.\ (\ref{eq:prodrule}), the simplest way to restore Hermiticity in the evaluation of $\nabla \cdot \left(B \nabla \psi \right)$ is to compute two consecutive first-order derivatives. 
But, as discussed in Sec.\ \ref{sec:num_grid},  this creates a problem with   the second derivative  that involves the Fourier component $\widetilde{\psi}(k_{N_x/2+1})$.
According to Ref.\ \cite{Johnson2011}, one should keep the term $\widetilde{\psi}(k_{N_x/2+1})$  in the two first derivatives, 
and average the results of $k_{N_x/2+1}= \pm \frac{N_x}{2} \delta k$ to maintain the symmetry in Fourier space. 
One can show that this ``average'' algorithm is equivalent to Algorithm \ref{algorithm} (Algorithm 3 in \cite{Johnson2011}), 
which is simpler to compute and thus implemented in {\sc HFBFFT}. In Algorithm \ref{algorithm}, one first computes an FFT-based first derivative, with $\widetilde{\psi}(k_{N_x/2+1})$ saved and then zeroed before the inverse FFT is performed on $\mathrm{i}k_n\widetilde{\psi}(k_n)$ (steps 1 through 3).
Then one multiplies in coordinate space with the field $B(x)$ involved, i.e., $\phi=B\psi^\prime$ (step 4). Finally, one computes the derivative of the $\phi$ with $\widetilde{\phi}(k_{N_x/2+1})$ modified so that we can keep Hermiticity without losing the information of $\widetilde{\psi}(k_{N_x/2+1})$ (steps 5 through 7).
The position-varying differential operator also appears in many other physics equations, like the heat equation with varying diffusivity and Poisson’s equation with changing permittivity;
hence, Algorithm \ref{algorithm} has a broad application range.

\begin{algorithm}
\setstretch{1.5}
\caption{Compute the one-dimensional position-varing differetiation  $\frac{d}{dx}\left[B(x)\frac{d\psi}{dx}\right]$.}
\label{algorithm}
\begin{algorithmic}[1]
\State  Compute Fourier transform
$\widetilde{\psi}_n=\mathtt{FFT}[\psi_\nu]$
       with $\psi_\nu=\psi(x_\nu)$.
\State Save $\widetilde{\psi}_{N_x/2+1}\rightarrow\widetilde{\Psi}$, build
       $\widetilde{\psi^\prime}_n=\mathrm{i}k_n\widetilde{\psi}_n$ with $\widetilde{\psi^\prime}_{N_x/2+1}=0$.
\State Compute inverse transform
       $\psi^\prime_\nu=\mathtt{FFT}^{-1}[\widetilde{\psi^\prime}_n]$.
\State  Build $\phi_\nu=B_\nu\psi^\prime_\nu$ with $B_\nu=B(x_\nu)$.
\State  Compute Fourier transform $\widetilde{\phi}_n=\mathtt{FFT}[\phi_\nu]$.
\State  Build $\widetilde{\phi^\prime}_n=\mathrm{i}k_n\widetilde{\phi}_n$ and set
        $\widetilde{\phi^\prime}_{N_x/2+1} = -\frac{\sum_{\nu=1}^{N_x}B_\nu}{N_x} \left(\frac{N_x}{2} \delta k \right)^2\widetilde{\Psi}$.
\State Compute inverse transform 
       $\frac{d}{dx}\left[B(x)\frac{d\psi}{dx}\right]_\nu=
        \mathtt{FFT}^{-1}[\widetilde{\phi^\prime}_n]$.
\end{algorithmic}
\end{algorithm}

\subsection{Numerical realization in harmonic-oscillator basis}\label{sec:ho_representation}

The HFB solutions obtained with the {\sc HFBFFT} code will be compared with the  well-established code {\sc HFBTHO}. This code has been extensively documented in several publications \cite{Stoitsov2005, Perez2017}. 
The solver
{\sc HFBTHO} uses an expansion of the s.p.\ wave functions in the basis of axially symmetric  HO (or transformed HO) states. The basis is given by the number of oscillator shells that defines the s.p.\ space size, as well as the oscillator length and deformation that determine the HO wave functions.
Local fields in {\sc HFBTHO} are handled on the Gaussian integration points and the
Gaussian integration rule is used to compute integrals.

A major difference between the two codes lies in the way the HFB
equations are solved. {\sc HFBFFT} uses a representation in terms of the canonical basis, see Sec.\ \ref{sec:algorithm}, while {\sc HFBTHO} works in
a quasiparticle space. 
The results are fully equivalent if the same number of s.p.\ states is used. 
Differences appear in connection with the cutoff in pairing space. {\sc HFBFFT} defines the cutoff in terms of the canonical s.p.\ energies, whereas {\sc HFBTHO} does that in terms of the quasiparticle energies.
This, taken together with the fact that the pairing strength has to depend on the size of the pairing space, means that the values of $V_{\mathrm{pair},q}$ are not fully portable. 
It will play a role in the benchmarking tests presented in Sec.\ \ref{sec:benchmark}.

\section{Benchmarks}\label{sec:benchmark}

In this section, we benchmark  {\sc HFBFFT}  against {\sc HFBTHO}, {\sc Sky1D}, and {\sc Sky2D}. 
These codes have symmetry restrictions. 
{\sc Sky1D} enforces spherical symmetry and can be used for magic nuclei.  
{\sc HFBTHO} and {\sc Sky2D} allow for axially symmetric shapes and cover all test cases here. 
Those codes can run with or without imposing reflection symmetry. 

First, we determine appropriate parameters to use, including the box size and grid spacing.
Before making comparisons with other solvers, we quantify the effect brought by the Hermiticity restoration.
In the next step, we compare some characteristic nuclei ranging from spherical doubly magic  $^{132}$Sn and $^{208}$Pb, to spherical superfluid $^{120}$Sn, to deformed superfluid $^{102,110}$Zr,  to superdeformed fission isomer in $^{240}$Pu.
In all these calculations, we use the Skyrme functional SLy4 \cite{Chabanat1998} in the particle-hole channel and the mixed density-dependent
$\delta$ interaction ($\rho_{0,\mathrm{pair}}=0.32$ fm$^{-3}$ in Eq.\ (\ref{eq:epair})) in the particle-particle channel.

\subsection{Parameter determination}
To ensure the correct asymptotic behavior near the box boundary, we use $^{110}$Zr to determine the appropriate box and grid sizes.
The nucleus $^{110}$Zr is chosen because it has a significant neutron excess and thus weakly bound canonical states.
The calculated proton and neutron densities were inspected for different box lengths and different grids. 
Based on this analysis, we adopted a cubic box with a side length of 37.6 fm and 48 grid points in each dimension (spacing between two neighboring points is 0.8 fm).
With the above settings, the proton and neutron densities are below 10$^{-7}$ nucleons/fm$^3$ at the boundary, which is small enough for our tests.
For spherical nuclei such as $^{120}$Sn, a smaller box is usually sufficient.

We take 176 neutron and 126 proton canonical states ($\Omega_\mathrm{n} = 176,\ \Omega_\mathrm{p} = 126$), 15 MeV energy cutoff for the pairing window.
This number of active states is determined by the tests for spherical $^{120}$Sn and deformed $^{110}$Zr nuclei.
When we increase the number of active states to 200 neutron and 150 proton states, the total energy remains stable within 10 keV.
In order to speed up the convergence, we perform 100 sub-iteration steps in the configuration space between two gradient iterations in the coordinate space, initialize with 30 HF+BCS steps, and employ the pairing enhancement factors defined in Sec.\ \ref{sec:breakdown}.

For {\sc HFBTHO} calculations, we take 25 HO shells  for both protons and neutrons unless explicitly stated otherwise.
An axially deformed HO basis with $\beta_2 = 0.2$ is used in deformed ground-state calculations ($^{102,110}$Zr and $^{240}$Pu) and $\beta_2 = 0.6$ is used to calculate the $^{240}$Pu fission isomer.
For the spherical nuclei, we also compare {\sc HFBFFT}  with the results  of the 1D spherical HFB code {\sc Sky1D}, which  uses a radial coordinate-space mesh and the five-point finite difference formula for derivatives. The mesh spacing and the number of points we employ in {\sc Sky1D} are 0.15 fm and 141, respectively.
For the deformed nuclei, we compare {\sc HFBFFT} results with the 2D axial HFB code {\sc Sky2D}, which uses 31 points in both $r$- and $z$-directions with  a mesh spacing of 0.7 fm. Since the nuclei considered in this study are all reflection-symmetric, the grid extends from $z=0$\,fm to $z=21$\,fm.

\subsection{Pairing renormalization}\label{sec:renormalization}
As we mentioned in Sec.\ \ref{sec:ho_representation}, pairing strengths are not portable between {\sc HFBFFT} and {\sc HFBTHO} because of different descriptions of the pairing space and different structure of the one-quasiparticle continuum in these two solvers. 
Therefore, we need to renormalize the pairing strengths to compare results for open-shell nuclei in which pairing is essential.
Intuitively, there are several choices for pairing renormalization.
 
For instance, one can tune pairing strengths to reproduce the pairing energies in different solvers.
However, as discussed in \cite{Papenbrock1999,Borycki2006}, the pairing energy density is divergent with respect to the cutoff energy. A better measure is the  quantity
\begin{equation}\label{eq:Ekineff}
\tilde{E}_\mathrm{kin}^q = E_\mathrm{kin}^q+ E_\mathrm{pair}^q~~(q={\rm n\,or\,p}),
\end{equation}
which is less sensitive to the pairing  cutoff energy.
As it will be
shown in Sec.\ \ref{sec:doublymagic},  the kinetic energy strongly depends on the  basis size in {\sc HFBTHO}.
Therefore, in situations when the error related to the choice of the basis, or spatial grid, dominates, $\tilde{E}_\mathrm{kin}$ will be a poor renormalization measure.
Another pairing measure is the spectral pairing gap \cite{Dobaczewski1984,Dobaczewski1996,Bender2000}
\begin{equation}\label{eq:Gap}
\Delta^q \equiv  \frac{\sum_{\alpha \in q} w_\alpha  v_\alpha^2 \Delta_{\alpha\alpha}}{\sum_{\alpha \in q} w_\alpha v_\alpha^2}
~~(q={\rm n\,or\,p}).
\end{equation}
This quantity has been used in numerous papers to adjust pairing strengths to observed odd-even mass differences and we shall use it in this study to renormalize the pairing channel of different solvers.

\subsection{Energy shift by Hermiticity restoration} \label{sec:hermiticityresult}

As we mentioned in Sec.\ \ref{sec:hermiticity}, the product rule in the FFT-based differentiation violates the Hermiticity of the position-varying differential operator.
To restore the Hermiticity, we implement Algorithm \ref{algorithm} in {\sc HFBFFT}.
The results are shown in Table \ref{tab:hermiticity} for several nuclei.
The Hermiticity violation is demonstrated  by a non-vanishing $\Delta\mathcal{S} \sim 10^{-7}$ MeV
in the calculations of spherical nuclei $^{132}$Sn and $^{208}$Pb where the static pairing vanishes and hence the HFB calculation is reduced  to  HF.
As for other open-shell nuclei with non-vanishing pairing, their $\Delta\mathcal{S}$ values are similar before and after the Hermiticity restoration.
These values of $\Delta\mathcal{S}$  are characteristic of the accuracy typically achieved in {\sc HFBFFT} and they are larger than the error due to the Hermiticity breaking.
In terms of the total energy,  the effect is of the order of a few keV, i.e., insignificant for many practical applications. 
Even so, Hermiticity breaking effects can affect some calculations if not remedied.
For example, the small error brought by the Hermiticity breaking can accumulate step by step in a time-dependent calculation.

\begin{table}[htb]
\centering
\begin{tabular}{c|rr|rr}
\hline
\hline
\multirow{2}{*}{{\sc HFBFFT}}  & \multicolumn{2}{c|}{Hermiticity broken} & \multicolumn{2}{c}{Hermiticity restored} \\
    & \multicolumn{1}{c}{$E_\mathrm{tot}$} &\multicolumn{1}{c|}{$\Delta\mathcal{S}$}          & \multicolumn{1}{c}{$E_\mathrm{tot}$} &\multicolumn{1}{c}{$\Delta\mathcal{S}$} \\
\hline 
$^{132}$Sn  &$-$1103.542\textbf{9} &3.44E-07  &$-$1103.542\textbf{3}  &1.60E-15 \\
$^{208}$Pb  &$-$1635.68\textbf{17} &3.16E-07 &$-$1635.68\textbf{07}   &1.20E-15\\
$^{120}$Sn  &$-$1018.331\textbf{0} &3.44E-07       &$-$1018.330\textbf{5}  &4.01E-07  \\
$^{110}$Zr  &$-$893.857\textbf{8} &4.59E-07      &$-$893.857\textbf{4}  &5.54E-07     \\
$^{102}$Zr  &$-$859.469\textbf{6}  &4.94E-07            &$-$859.469\textbf{2}  &3.93E-07       \\
\hline
\hline
\end{tabular}
\caption{Total energies $E_\mathrm{tot}$ (in MeV) and $\Delta\mathcal{S}$ (in MeV) for five nuclei calculated with  {\sc HFBFFT} without and with the Hermiticity restoration. The digits which do not coincide before and after the Hermiticity restoration are marked in bold.}
\label{tab:hermiticity}
\end{table}

\subsection{Doubly magic nuclei: $^{132}$Sn and $^{208}$Pb}\label{sec:doublymagic}
In the first step, we calculate two doubly magic unpaired  nuclei $^{132}$Sn and $^{208}$Pb.
For these nuclei, the results of \mbox{\sc HFBFFT} and {\sc Sky3D} are identical.
In Table \ref{tab:132Sn}, we list the  ground-state energies as well as contributions from various functional terms,  obtained from four solvers {\sc HFBFFT}, {\sc HFBTHO}, {\sc Sky1D} and {\sc Sky2D} for $^{132}$Sn. 
Table~\ref{tab:208Pb} shows similar results for $^{208}$Pb.
\begin{table}[htb]
\centering
\begin{tabular}{lrrrr}
\hline
\hline
$^{132}$Sn                    & {\sc HFBTHO}     & {\sc HFBFFT}  & {\sc Sky1D} & {\sc Sky2D}\\
\hline
$E_{\mathrm{tot}}$                  & $-$1103.4\textbf{9}	& $-$1103.5\textbf{4} & $-$1103.5\textbf{7}	& $-$1103.5\textbf{6}\\
$E_{\mathrm{kin}}^\mathrm{n}$                & 1637.\textbf{71}	& 1637.9\textbf{7}	& 1638.0\textbf{1} & 1638.0\textbf{2}\\
$E_{\mathrm{kin}}^\mathrm{p}$                & 808.\textbf{44}	& 808.5\textbf{7}	& 808.5\textbf{9} &808.5\textbf{6}\\
$E_{\mathrm{\rho \rho}}$                & $-$487\textbf{6.26}	&  $-$4877.0\textbf{2} & $-$4877.0\textbf{4}	&$-$4877.0\textbf{7} \\
$E_{\mathrm{\rho \tau}}$                & 821.\textbf{49}	& 821.7\textbf{0}	& 821.7\textbf{3}  &821.7\textbf{2}\\
$E_{\mathrm{\rho \Delta \rho}}$                & 248.\textbf{11}	 & 248.2\textbf{3}	& 248.2\textbf{5} &248.2\textbf{3}\\
$E_{\mathrm{\rho \nabla \vec{J}}}$                & $-$84.4\textbf{0}	& $-$84.4\textbf{3} & $-$84.4\textbf{4}	 &$-$84.4\textbf{3}\\
$E_{\mathrm{Coul}}$                & 341.4\textbf{2} & 341.4\textbf{4} & 341.4\textbf{4} &341.4\textbf{3}	\\
\hline
\hline
\end{tabular}
\caption{Energy contributions (in MeV) to the binding energy  for $^{132}$Sn computed with {\sc HFBTHO}, {\sc HFBFFT}, {\sc Sky1D},  and {\sc Sky2D}. The digits which do not coincide with {\sc HFBFFT} are marked in bold.}
\label{tab:132Sn}
\end{table}
When we compare {\sc HFBFFT} with {\sc Sky1D} and {\sc Sky2D} for $^{132}$Sn , we find the  energy differences do not usually exceed 40\,keV. 
Such small differences  can be traced back to   different box boundary conditions assumed in these codes.
In {\sc HFBFFT}, calculations are performed in a 3D rectangular box while  the box is represented by  a spherical shell in {\sc Sky1D} and a cylindrical shape in {\sc Sky2D}. 
For a well-bound nucleus and large spatial boxes, the results should be practically independent of the geometry of the box. 
As seen in Table\,\ref{tab:132Sn} this indeed holds for $^{132}$Sn. 
As we will see below, larger box-related errors are expected in superfluid and/or weakly bound nuclei. 
For nuclear matter and time-dependent calculations, the finite-size box errors can be appreciable; they can be greatly reduced by imposing twist-averaged boundary conditions
\cite{Schuetrumpf2016}.

\begin{table*}[htb]
\centering
\begin{tabular}{lrrrrrr}
\hline
\hline
$^{208}$Pb                    & $N$=15     & $N$=20 & $N$=25 & $N$=30 & {\sc HFBFFT}  & {\sc Sky1D} \\
\hline
$E_{\mathrm{tot}}$                 & $-$163\textbf{4.25}	 & $-$1635.\textbf{16}	 & $-$1635.\textbf{46}	 & $-$1635.6\textbf{2} &$-$1635.6\textbf{8}	& $-$1635.7\textbf{0} \\
$E_{\mathrm{kin}}^\mathrm{n}$               & 252\textbf{5.13}	& 252\textbf{7.80}	& 252\textbf{8.42}	& 2528.\textbf{83} &2529.1\textbf{3} & 2529.1\textbf{6}	\\
$E_{\mathrm{kin}}^\mathrm{p}$              & 133\textbf{4.56}	& 133\textbf{6.34}	& 1336.\textbf{71}	& 1336.\textbf{91}  &1337.0\textbf{6}	& 1337.0\textbf{7}  \\
$E_{\mathrm{\rho \rho}}$               & $-$78\textbf{35.80}	& $-$784\textbf{4.07}	& $-$784\textbf{5.66}	 & $-$784\textbf{6.67}  & $-$7847.\textbf{54}& $-$7847.\textbf{63} \\
$E_{\mathrm{\rho \tau}}$                & 132\textbf{7.84}	& 132\textbf{9.55}	& 1329.\textbf{79}	& 1329.\textbf{98} &1330.2\textbf{0}	& 1330.2\textbf{2}  \\
$E_{\mathrm{\rho \Delta \rho}}$               & 31\textbf{4.05} &  315.\textbf{12}	& 315.\textbf{12} 	& 315.\textbf{17}  &315.2\textbf{9}	& 315.2\textbf{9} \\
$E_{\mathrm{\rho \nabla \vec{J}}}$                & $-$96.\textbf{30}	& $-$96.4\textbf{4}	& $-$96.4\textbf{2} 	& $-$96.4\textbf{3} &$-$96.4\textbf{5} & $-$96.4\textbf{5}	\\
$E_{\mathrm{Coul}}$                & 796.\textbf{26} & 796.\textbf{55}	& 796.\textbf{56}	& 796.6\textbf{0} &796.6\textbf{3} & 796.6\textbf{3}	 \\
\hline
\hline
\end{tabular}
\caption{Energies for $^{208}$Pb from {\sc HFBTHO} (computed with different numbers of HO shells $N$), {\sc HFBFFT}, and {\sc Sky1D}. All energies are in MeV. The digits which do not coincide with {\sc HFBFFT} are marked in bold.}
\label{tab:208Pb}
\end{table*}

We find about 50\,keV energy difference between {\sc HFBTHO} and {\sc HFBFFT};  this difference can be primarily traced back to $E_\mathrm{kin}$ and  $E_\mathrm{\rho \rho}$.
As discussed in Refs.\ \cite{Furnstahl2012, Binder2016}, the kinetic energy converges slowly in the HO basis. 
To investigate this effect, we calculate $^{208}$Pb using different numbers of HO\ shells in {\sc HFBTHO}.
We see in Table \ref{tab:208Pb}  that when we increase the number of HO shells to 30, the {\sc HFBTHO} energies approach the {\sc HFBFFT} values.
It is also seen that  $E_\mathrm{kin}$ and $E_\mathrm{\rho \rho}$ exhibit the largest variations with $N$.

\begin{figure}[htb]
\centering
\includegraphics[width=0.45\textwidth]{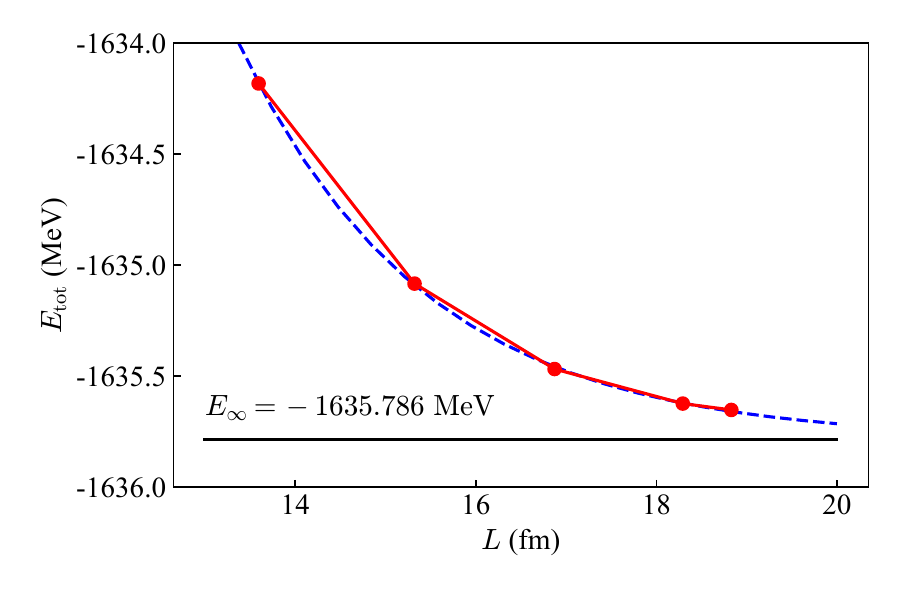}
\caption{$E_\mathrm{tot}$ as a function of $L$  for $^{208}$Pb. The  {\sc HFBTHO} results are marked by red dots. The blue curve is fitted according to  Eq.\ (\ref{eq:ene_conv}).}
\label{fig:1}
\end{figure}

In Refs.\ \cite{Furnstahl2012, More2013}, the correction to the g.s.\ energy due to the finite number of HO shells $N$ has been derived:
\begin{equation}
	\label{eq:ene_conv}
	E_{L} =  E_{\infty} + a_0e^{-2k_{\infty}L},
\end{equation}
where  $L \equiv \sqrt{2(N+3/2+2)}b$, $b$ is the oscillator length of our HO basis, and $a_0,k_{\infty}$ and $E_{\infty}$ are fit parameters. 
Then $E_{\infty}$ is  the energy in the limit of infinitely large model space. The fit 
of $E_\mathrm{tot}$  to Eq.\ (\ref{eq:ene_conv}) 
is presented in Fig.\ \ref{fig:1} and the resulting value of  $E_{\infty} =  -1635.786$ MeV  agrees fairly well with the  {\sc HFBFFT} and {\sc Sky1D} values.
Hence, obtaining an accurate kinetic as well as total energies in a HO basis-expansion solver requires a huge number of shells. In this context, the use of the coordinate-space representation is beneficial.

\begin{table*}[htb]
\centering
\begin{tabular}{l|r|rrr|rrr}
\hline
\hline
\multirow{2}{*}{$^{120}$Sn}    & {\sc HFBTHO}    & {\sc  HFBFFT}    & {\sc Sky1D} & {\sc Sky2D} & {\sc HFBFFT }  & {\sc Sky1D}  & {\sc  Sky2D} \\ 
 &    & \multicolumn{3}{c}{$\tilde{E}_\mathrm{kin}^\mathrm{n}$-renorm. }  &
  \multicolumn{3}{|c}{$\Delta^\mathrm{n}$ -renorm. } \\
  & & & & & & & \\[-8pt]
\hline
$E_{\mathrm{tot}}$                    & $-$1018.\textbf{77} &$-$1018.3\textbf{4} &$-$1018.\textbf{45} &$-$1018.3\textbf{7} &$-$1018.7\textbf{8} &$-$1018.\textbf{92} &$-$1018.7\textbf{4}\\
$E_{\mathrm{Coul}}$                   & 347.\textbf{37}   & 347.4\textbf{4}  & 347.4\textbf{5} &347.4\textbf{1} &347.4\textbf{7}  & 347.4\textbf{9} &347.4\textbf{5}\\
$E_{\mathrm{kin}}^\mathrm{n}$  & 134\textbf{0.51}  & 1335.4\textbf{0}  &1335.\textbf{18} &1335.4\textbf{3} &1339.1\textbf{7}  &1339.1\textbf{4} &1338.\textbf{72}\\
$E_{\mathrm{kin}}^\mathrm{p}$  & 830.\textbf{75}    & 830.9\textbf{7}   &831.0\textbf{1} &830.0\textbf{1} &831.2\textbf{5}  &831.\textbf{31} &831.2\textbf{8}\\
$E_\mathrm{pair}^\mathrm{n}$                & $-$\textbf{12.48}   & $-$7.3\textbf{7}   &$-$7.\textbf{15} &$-$7.4\textbf{0} & $-$9.\textbf{29} &$-$9.\textbf{14} &$-$9.\textbf{02}\\
$\tilde{E}_{\mathrm{kin}}^\mathrm{n}$ & \textit{1328.03}  & \textit{1328.03}  &\textit{1328.03}
&\textit{1328.03} &1329.88 &1330.01 &1329.70\\
$\Delta^\mathrm{n}$                  & \textit{1.25}     & 1.08     &1.07 &1.09 &\textit{1.25} &\textit{1.25} &\textit{1.25}\\
$\epsilon_{\mathrm{F},\mathrm{n}}$                 & $-$8.0\textbf{2}    &$-$8.0\textbf{1}     &$-$8.0\textbf{1} &$-$8.0\textbf{4} &$-$8.0\textbf{0}  &$-$8.0\textbf{0} &$-$8.0\textbf{4}\\
$V_\mathrm{pair,n}$                      & $-$284.57   & $-$342.70   &$-$346.50  &$-$354.90 & $-$361.80 & $-$367.30 &$-$372.35\\ 
$r_\mathrm{rms}$                  & 4.67     & 4.67   &4.67 &4.67 & 4.67 &4.67 &4.67\\

\hline
\hline  
\end{tabular}
\caption{Results of HFB + SLy4 calculations for $^{120}$Sn using {\sc HFBTHO}, {\sc HFBFFT}, {\sc Sky1D}, and  {\sc Sky2D}.
Two neutron pairing renormalization variants are considered, by adjusting 
the neutron pairing strengths in {\sc HFBFFT}, {\sc Sky1D}, and  {\sc Sky2D}  to reproduce the {\sc HFBTHO} values of $\tilde{E}_\mathrm{kin}^\mathrm{n}$ and $\Delta^\mathrm{n}$.
 All energies are in  MeV. 
 The radius $r_{\mathrm{rms}}$ is in fm.
 The digits which do not coincide with {\sc HFBFFT} are marked in bold.}
\label{tab:120Sn}
\end{table*}

\subsection{Spherical superfluid nucleus: $^{120}$Sn}\label{sec:120Sn}

We now calculate $^{120}$Sn which has a non-vanishing neutron pairing.
The neutron pairing strength $V_\mathrm{pair,n}$ in {\sc HFBTHO} is adjusted to the average experimental neutron pairing gap $\Delta^\mathrm{n} = 1.25$\,MeV.
In {\sc HFBFFT}, {\sc Sky1D} and {\sc Sky2D}, two pairing renormalizations are used.
In the first variant, the neutron pairing strengths are adjusted to reproduce the {\sc HFBTHO} value of $\tilde{E}_\mathrm{kin}^\mathrm{n}$.
In the second variant, the {\sc HFBTHO} value of $\Delta^\mathrm{n}$ is matched.
The results for both variants are displayed in Table \ref{tab:120Sn}.
The neutron pairing strengths vary between the solvers, reflecting different structure of their quasiparticle pairing spaces, i.e., different pairing cutoff procedures and different structure of the discretized one-quasiparticle continuum.

Although there are large discrepancies in $E_\mathrm{kin}^\mathrm{n}$ and $E_\mathrm{pair}^\mathrm{n}$ between {\sc HFBFFT} and {\sc HFBTHO}, in the first renormalization variant, the difference of the total energy, about 0.4 MeV, is quite reasonable considering the fact that the pairing space is treated differently and the {\sc HFBTHO} results are affected by the basis truncation error.
The difference in $E_{\mathrm{tot}}$ between the three coordinate-space solvers, less than 150\,keV,  reflects the dependence  of the level density of the discretized quasiparticle continuum on the box boundary conditions assumed.

In the pairing gap renormalization variant, the agreement of $E_{\mathrm{tot}}$ is even better, with only 10-30\,keV difference between {\sc HFBFFT}, {\sc HFBTHO} and {\sc Sky2D}. 
In this variant,  the magnitudes of the neutron pairing energy and kinetic energy are considerably larger as compared to the variant in which $\tilde{E}_\mathrm{kin}^\mathrm{n}$ is renormalized.
Still, as seen  in Table \ref{tab:120Sn},  both  pairing renormalizations work reasonably well for $^{120}$Sn.  
It is interesting to note that the total root-mean-square (rms)  radii $r_\mathrm{rms}$   are predicted very robustly in all renormalization variants.

\subsection{Axially deformed nuclei: $^{102,110}$Zr}

The neutron-rich nuclei $^{102,110}$Zr are suitable test cases, as they are known/expected to have large prolate deformations. 
In addition, $^{110}$Zr is weakly bound, with the neutron chemical potential $\epsilon_{\mathrm{F},\mathrm{n}}\approx -3.5$\,MeV.  
The HFB proton pairing vanishes in this nucleus.
In Table \ref{tab:110Zr}, we show results for $^{110}$Zr with the two pairing renormalization schemes investigated in Sec.~\ref{sec:120Sn}.
It is seen that the {\sc HFBFFT} results  for various observables, i.e., total energy, quadrupole moments, and the rms  radius, all agree well with those from  {\sc HFBTHO} in both pairing variants.

 \begin{table*}[htb]
\centering
\begin{tabular}{l|r|rr|rr}
\hline
\hline
\multirow{2}{*}{$^{110}$Zr }    & {\sc HFBTHO}    & {\sc  HFBFFT}     & {\sc Sky2D} & {\sc HFBFFT }   & {\sc  Sky2D} \\ 
 &    & \multicolumn{2}{c}{$\tilde{E}_\mathrm{kin}^\mathrm{n}$-renorm. }  &
  \multicolumn{2}{|c}{$\Delta^\mathrm{n}$ -renorm. } \\
 \hline
$E_\mathrm{tot}$                     & $-$893.\textbf{97}  & $-$894.3\textbf{3} &$-$894.3\textbf{2} &$-$894.0\textbf{1} &$-$894.0\textbf{1}\\
$E_\mathrm{Coul}$                  & 226.7\textbf{2}   & 226.7\textbf{2} &226.7\textbf{1} & 226.7\textbf{4} &226.7\textbf{0}\\
$E_\mathrm{kin}^\mathrm{n}$  & 136\textbf{8.08}   & 1369.\textbf{22} &1368.\textbf{98} & 1367.\textbf{86}  &1367.\textbf{13}\\
$E_\mathrm{kin}^\mathrm{p}$  & 632.0\textbf{3}  & 632.0\textbf{5} &632.\textbf{13}  & 632.\textbf{16}  &632.\textbf{05}\\
$E_\mathrm{pair}^\mathrm{n}$                & $-$\textbf{3.18}   &$-$4.\textbf{31} &$-$4.\textbf{08} & $-$2.\textbf{30} &$-$2.\textbf{19}\\
$\tilde{E}_\mathrm{kin}^\mathrm{n}$ & \textit{1364.90}  & \textit{1364.90} &\textit{1364.90} &1365.56 &1364.94\\
$\Delta^\mathrm{n}$                  & \textit{0.64}    & 0.93 &0.92 &\textit{0.64} &\textit{0.64} \\
$\epsilon_{\mathrm{F},\mathrm{n}}$                 & $-$3.5\textbf{5}   & $-$3.5\textbf{0} &$-$3.5\textbf{2} &$-$3.5\textbf{5} &$-$3.5\textbf{7}\\
$V_\mathrm{pair,n}$                      & $-$284.57  & $-$409.80 &$-$428.00 & $-$371.00 &$-$384.80 \\
$r_\mathrm{rms}$                          & 4.7\textbf{3}    & 4.7\textbf{3} &4.7\textbf{4} &4.7\textbf{3}  &4.7\textbf{4} \\
$Q_{20}^\mathrm{n}$  & 7\textbf{89}   & 79\textbf{4} &79\textbf{5} & 79\textbf{1} &79\textbf{6}\\
$Q_{20}^\mathrm{p}$  & 44\textbf{4}   & 44\textbf{7} &44\textbf{7} & 44\textbf{5} &44\textbf{7}\\

\hline
\hline 
\end{tabular}
\caption{Results of HFB + SLy4 calculations for $^{110}$Zr with {\sc HFBTHO}, {\sc HFBFFT} and {\sc Sky2D}.
Two neutron pairing renormalization variants are considered, by adjusting the neutron pairing strengths in {\sc HFBFFT} and {\sc Sky2D}  to reproduce the {\sc HFBTHO} values of $\tilde{E}_\mathrm{kin}^\mathrm{n}$ and $\Delta^\mathrm{n}$.
All energies are in MeV.
The radius $r_{\mathrm{rms}}$ is in fm and quadrupole moments $Q_{20}^\mathrm{p,n}$ are in fm$^2$. The HFB proton pairing vanishes in this nucleus.
The digits which do not coincide with {\sc HFBFFT} are marked in bold.}
\label{tab:110Zr}
\end{table*}

In the case of $^{102}$Zr, one also needs to consider proton pairing. 
In this case, we  renormalize both neutron and proton spectral pairing gaps by reproducing their values obtained from {\sc HFBTHO}.
In the calculation, 25 HO shells for both neutrons and protons are employed in {\sc HFBTHO}, which means that the s.p.\ proton and neutron spaces are the same.
In {\sc HFBFFT}, the canonical spaces are different as $\Omega_\mathrm{n} = 176$, $\Omega_\mathrm{p} = 126$. However,  the actual pairing space is set by the soft-cutoff  factor $w_\alpha$.
It is seen in Table \ref{tab:102Zr} that the benchmarking results following the pairing renormalization are very satisfactory. 
In particular, the results of {\sc HFBFFT}, {\sc HFBTHO}, and {\sc Sky2D }  are fairly close for the observables:
$E_\mathrm{tot}$, $r_\mathrm{rms}$ , and quadrupole moments.

\begin{table}[htb]
\centering
\begin{tabular}{lrrr}
\hline
\hline
$^{102}$Zr                    & HFBTHO   & HFBFFT  & Sky2D   \\
\hline
$E_\mathrm{tot}$                    & $-$859.6\textbf{5} & $-$859.6\textbf{9}  &$-$859.6\textbf{7}\\
$E_\mathrm{Coul}$                  & 231.1\textbf{1}  & 231.1\textbf{6}   &231.1\textbf{4}\\
$E_\mathrm{kin}^\mathrm{n}$  & 120\textbf{2.02} & 120\textbf{0.96} & 120\textbf{1.97}\\
$E_\mathrm{kin}^\mathrm{p}$  & 651.2\textbf{5}  & 651.2\textbf{2}  & 651.2\textbf{7}\\
$E_\mathrm{pair}^\mathrm{n}$               & $-$\textbf{3.39}   & $-$2.\textbf{50}   &$-$2.\textbf{39} \\
$E_\mathrm{pair}^\mathrm{p}$              & $-$1.\textbf{97}   & $-$1.4\textbf{2}   &$-$1.3\textbf{8} \\
$\tilde{E}_\mathrm{kin}^\mathrm{n}$ & 119\textbf{8.63} & 1199.5\textbf{3} & 1199.5\textbf{8} \\
$\tilde{E}_\mathrm{kin}^\mathrm{p}$ & 649.\textbf{28}  & 649.\textbf{79} & 649.\textbf{89} \\
$\Delta^\mathrm{n}$                & \textit{0.69}   & \textit{0.69}   &\textit{0.69}\\
$\Delta^\mathrm{p}$                  & \textit{0.56}   & \textit{0.56}    & \textit{0.56} \\
$\epsilon_{\mathrm{F},\mathrm{n}}$                  & $-$5.4\textbf{3}   & $-$5.4\textbf{2}   & $-$5.4\textbf{4} \\
$\epsilon_{\mathrm{F},\mathrm{p}}$                & $-$12.0\textbf{9}  & $-$12.0\textbf{9}  &$-$12.1\textbf{0}  \\
$V_\mathrm{pair}^\mathrm{n}$                        & $-$284.57  & $-$367.00  & $-$378.40 \\
$V_\mathrm{pair}^\mathrm{p}$
  & $-$284.57  & $-$372.00  & $-$384.70 \\
$r_\mathrm{rms}$                         & 4.58    & 4.58     &4.58\\
$Q_{20}^\mathrm{n}$  & 63\textbf{9}  & 63\textbf{9} & 64\textbf{0}  \\
$Q_{20}^\mathrm{p}$  & 411  & 411  & 411 \\
\hline
\hline
\end{tabular}
\caption{Results of HFB + SLy4 calculations for $^{102}$Zr using {\sc HFBTHO}, {\sc HFBFFT} and {\sc Sky2D}. The pairing renormalization is carried out by adjusting the proton and neutron  pairing strengths in {\sc HFBFFT} and {\sc Sky2D}  to reproduce the {\sc HFBTHO} values of $\Delta^\mathrm{n}$ and $\Delta^\mathrm{p}$.
All energies are in MeV. 
The radius $r_{\mathrm{rms}}$ is in fm and quadrupole moments $Q_{20}^\mathrm{p,n}$ are in fm$^2$.
The digits which do not coincide with {\sc HFBFFT} are marked in bold.}
\label{tab:102Zr}
\end{table}

\subsection{Superdeformed  heavy nucleus: $^{240}$Pu}
Compared with the HO basis, the coordinate-space representation can better capture strongly deformed configurations, such as
the superdeformed fission isomer (f.i.) of $^{240}$Pu. 
Indeed, very large configuration spaces are needed to guarantee the convergence of the HO expansion at  large deformations \cite{Nikolov2011,Schunck2013}.
Given the large number of  nucleons in $^{240}$Pu, one needs to carefully consider the number of canonical states in  {\sc HFBFFT} and {\sc Sky2D} calculations.
To this end, we performed a series of calculations by increasing the canonical space until the convergence had been reached.
This has been done separately for the ground state (g.s.) and f.i.\ of $^{240}$Pu.
The final values are: ($\Omega_\mathrm{n},\   \Omega_\mathrm{p}) = (300,\ 200)$ for the g.s.\ and ($\Omega_\mathrm{n},\  \Omega_\mathrm{p}) = (400,\ 300)$ for the f.i.\ calculations.
We renormalize the pairing strengths for $^{240}$Pu to reproduce the g.s.\  $\Delta^{\mathrm{n}}$ and $\Delta^{\mathrm{p}}$ obtained in {\sc HFBTHO}.
The results are displayed in Table \ref{tab:240Pu}. 

In {\sc HFBTHO} and {\sc Sky2D}, the f.i.\ is found by performing  quadrupole-moment constrained calculations.
The f.i.\ configuration  in {\sc HFBFFT} was computed  by initializing the code with various HO deformations.
As seen in  Table \ref{tab:240Pu}, {\sc HFBTHO} and {\sc HFBFFT} results are very similar for  g.s.\ energies,  g.s.\ quadruple deformations, and radii.

To test the functionality of {\sc HFBFFT}  for the f.i.,
we renormalize pairing strengths in {\sc HFBFFT} and {\sc Sky2D}  to the {\sc HFBTHO} pairing gaps.
Both coordinate-space solvers give very close results for the f.i., and they agree nicely with the {\sc HFBTHO} results, see Table \ref{tab:240Pu}.
Overall, the $^{240}$Pu results obtained with {\sc HFBFFT} for both the g.s.\ and f.i.\ show reasonable agreement with those from {\sc HFBTHO}.

\begin{table*}[htp]
\centering
\begin{tabular}{l|rr|rrr}
\hline
\hline
\multirow{3}{*}{$^{240}$Pu}  & \multicolumn{2}{c|}{ground state} & \multicolumn{3}{c}{fission isomer} \\ 

    & \multicolumn{1}{c}{{\sc HFBTHO}}  &\multicolumn{1}{c|}{{\sc HFBFFT}}         &\multicolumn{1}{c}{{\sc HFBTHO}}  
    &\multicolumn{1}{c}{{\sc HFBFFT}} 
    &\multicolumn{1}{c}{{\sc Sky2D}}\\
\hline
$E_\mathrm{tot}$      & $-$1802.\textbf{11}  &$-$1802.\textbf{43} & $-$1797.\textbf{00}   &$-$1797.\textbf{35} &$-$1797.3\textbf{5}\\
$E_\mathrm{Coul}$      & 9\textbf{89.61}     &956.\textbf{98}  & 957.0\textbf{2}       &  956.9\textbf{6} & 956.9\textbf{0}\\
$E_\mathrm{kin}^\mathrm{n}$      & 293\textbf{8.92}     &293\textbf{9.94} & 292\textbf{2.56}   &  2923.4\textbf{5} &2923.4\textbf{3} \\
$E_\mathrm{kin}^\mathrm{p}$     & 152\textbf{0.95}    &152\textbf{1.46} & 1525.\textbf{25}     &  1525.\textbf{52} & 1525.\textbf{33}\\
$E_\mathrm{pair}^\mathrm{n}$       & $-$\textbf{3.11}       &$-$\textbf{2.30} & $-$\textbf{3.52}   &  $-$2.\textbf{60} & $-$2.\textbf{48}\\
$E_\mathrm{pair}^\mathrm{p}$       & $-$1.\textbf{54}          &$-$1.\textbf{22} & $-$2.\textbf{85}      &$-$2.\textbf{19} &$-$2.\textbf{07}\\
$\tilde{E}_\mathrm{kin}^\mathrm{n}$     & 293\textbf{5.81}     & 293\textbf{7.64} & 291\textbf{9.03}   & 2920.\textbf{85} &2920.\textbf{55}\\
$\tilde{E}_\mathrm{kin}^\mathrm{p}$      & 151\textbf{9.40}     &152\textbf{0.25} & 152\textbf{2.39}   &1523.\textbf{33} &1523.\textbf{25} \\
$\Delta^\mathrm{n}$      &\textit{0.44}           &\textit{0.44} & \textit{0.47}  & \textit{0.47}  & \textit{0.47} \\
$\Delta^\mathrm{p}$      & \textit{0.33}          &\textit{0.33} & \textit{0.46}         &\textit{0.46} &\textit{0.46} \\
$\epsilon_{\mathrm{F},\mathrm{n}}$       & $-$5.7\textbf{1}      
& $-$5.7\textbf{0} &$-$5.6\textbf{6}        & $-$5.6\textbf{5} &$-$5.6\textbf{7}  \\
$\epsilon_{\mathrm{F},\mathrm{p}}$      & $-$5.6\textbf{9}       & $-$5.7\textbf{0}   & $-$5.7\textbf{6}    & $-$5.7\textbf{7} &$-$5.7\textbf{9}  \\
$r_\mathrm{rms}$         & 5.93          &5.93 & 6.40      & 6.40  &6.40   \\
$Q_{20}^\mathrm{n}$      & 178\textbf{4}    & 178\textbf{2} & 50\textbf{63}   & 507\textbf{2} &507\textbf{1}\\
$Q_{20}^\mathrm{p}$      & 116\textbf{6}     & 116\textbf{5} & 33\textbf{36} &  334\textbf{4} &334\textbf{3}\\
$V_\mathrm{pair}^\mathrm{n}$   & $-$284.57           &$-$360.00 & $-$284.57     &$-$369.00 &$-$384.60\\
$V_\mathrm{pair}^\mathrm{p}$   & $-$284.57           &$-$355.00  & $-$284.57     &$-$360.00  &$-$375.80\\
s.p. space     & 25 shells   & (300,\ 200) & 25 shells     & (400,\ 300) & (400,\ 300)\\
\hline
\hline  
\end{tabular}
\caption{Results of HFB + SLy4 calculations for $^{240}$Pu ground state and fission isomer using {\sc HFBTHO}, {\sc HFBFFT} and {\sc Sky2D}. 
The pairing strengths in {\sc HFBFFT} and {\sc Sky2D} were adjusted to reproduce the spectral pairing gaps obtained in {\sc HFBTHO} for the g.s.\ and f.i.\ separately.
The s.p.\ space for {\sc HFBFFT} is defined by means of ($\Omega_\mathrm{n}$, $\Omega_\mathrm{p}$). 
All energies are in  MeV, $r_{\mathrm{rms}}$ is in fm, and  $Q_{20}^\mathrm{p,n}$ are in fm$^2$.
The digits which do not coincide with {\sc HFBFFT} are marked in bold.}
\label{tab:240Pu}
\end{table*}

\section{Conclusions}\label{sec:conclusion}
We developed a 3D Skyrme HFB solver {\sc HFBFFT} in the coordinate-space representation using the canonical basis approach.
The code is based on the well-optimized {\sc Sky3D} solver.
In {\sc HFBFFT} we implemented several new elements to facilitate calculations, namely 
(i)  the sub-iteration method in configuration space to accelerate the convergence; 
(ii) the soft pairing cutoff and pairing annealing
to avoid  pairing breakdown; and (iii)
a new algorithm to restore the Hermiticity of the HFB matrix.

The new solver  has been  benchmarked
for several spherical and deformed nuclei against {\sc HFBTHO}, {\sc Sky2D}, and (for spherical systems)  {\sc Sky1D}. The representation of the  positive-energy continuum differs between HFB codes: In particular, it depends on the code's geometry (spherical, cylindrical, Cartesian),  the size of  s.p.\ configuration space (number of HO shells, box size, grid size), and the effective pairing space. Consequently,
even if the EDFs employed in two codes are identical, the pairing channel is usually described differently. This creates problems when comparing different HFB solvers as the perfect benchmarking is practically impossible \cite{Pei2008}. In this work, we carried our inter-code comparisons by renormalizing pairing strengths to the spectral  pairing gaps and/or the effective kinetic energy $\tilde{E}_\mathrm{kin}$. While both methods give similar results, spectral pairing gaps are less sensitive to the s.p.\ space assumed.

By carrying out calculations with different HFB solvers, we were able to assess the ranges of different uncertainties. 
For the total energy, the typical errors are: several keV due to the Hermiticity breaking; 10-80\,keV due to different box boundary conditions assumed;  10-140\,keV due to different quasiparticle continuum discretizations; and several hundred keV due to the basis truncation in HO basis-expansion solvers.

As a 3D solver, {\sc HFBFFT} is the tool of choice to  study deformed and weakly bound systems. 
To make this tool versatile,  several  enhancements are planned.
Most importantly,  we intend to  implement pairing regularization \cite{Bulgac02,Borycki2006,Pei2011} to get rid of the dependence of pairing strengths on the cutoff energy.
Another essential development is to be able to compute potential energy surfaces defined by means of constraining one-body operators. 
 This will enable us to use {\sc HFBFFT} in the calculations of  large-amplitude nuclear collective motions such as fission or fusion, for which the solvers based on the basis-expansion approach   require the use of excessively large configuration spaces.
Finally, the performance of {\sc HFBFFT} needs to be further optimized for modern supercomputer architectures.
Then the code will be ready for publication.

\section*{Acknowledgments}
Comments from Kyle Godbey are gratefully appreciated.
Computational resources were provided by the Institute for Cyber-Enabled Research at Michigan State University. 
This material is based upon work supported by the U.S.\ Department of Energy, Office of Science, Office of Nuclear Physics under award numbers DE-SC0013365 and DE-SC0018083 (NUCLEI SciDAC-4 collaboration).

\bibliographystyle{IEEEtran}

\bibliography{HFBFFT_ref}

\end{document}